\documentclass[11pt]{article}
\usepackage{lingmacros}
\usepackage{tree-dvips}
\usepackage{graphicx}
\usepackage{fullpage}
\usepackage{authblk}
\usepackage{algorithmicx}
\usepackage{algpseudocode}
\usepackage{algorithm}
\usepackage{amsthm}  
\usepackage{multirow}
\usepackage{array}

\date{January 20, 2016}
\author{ Samir Bouftass \\ E-mail : crypticator@gmail.com} 
\affil{Crypticator.Inc}

\title{\hspace{3 mm} On a new fast public key cryptosystem }

\begin{document}

\maketitle

\begin{abstract}

We analyze ModDiv2Inv, a problem consisting
on inverting the function below : \\*\\*
$F(X) =(A\times X)Mod(2^p)Div(2^q)$.\\*\\* Mod is modulo operation, Div is integer division operation, A, p and q are known integers where A is pseudorandom and p bits, X is q bits and $ p > 2 \times q $.\\*
We define p and q values for which ModDiv2Inv can be the hardest.\\*
We then present ModDiv2Kex, a new fast key exchange algorithm based on ModDiv2Inv .\\*

\end{abstract}


\section{Introduction :}
\vspace{4 mm}

\indent 
Since its invention by Withfield Diffie and Martin Hellman [1], public key cryptography has imposed
itself as the necessary and indispensable building block of every IT Security architecture.\\*
In the last decades, it has been proven that public key cryptosystems based on number theory problems are not immune againt quantum computing attacks [3], urging the necessity of inventing new algorithms not based on classical problems namely Factoring, Dicret log over multiplicative groups or elliptic curves.\\*
In this paper we analyze ModDiv2Inv a problem consisting on inverting the following function :\\*

 $F(x) =(A\times X)Mod(2^p)Div(2^q)$ .\\*
 
\noindent Mod is modulo operation, Div is integer division operation, A, p and q are known integers where A is pseudorandom and p bits, X is q bits and $ p > 2 \times q $.\\* 
We evaluate the hardness of this problem
by comparing it to Subset sum problem [4][5][6], one of the well studied NP complete problems.\\*

\noindent We present ModDiv2Kex, a new key exchange algorithm based on ModDiv2Inv.\\*

\noindent We introduce a computationnel assumption related to the hardest instances of ModDiv2Inv, and show what conditions ModDiv2Kex parameters should fullfill in order to be based on said assumption. We also show how ModDiv2Kex is efficient, compared to Diffie-Hellman[1] and RSA[2].

\section{Analysis : } 
\vspace{4 mm}

\subsection{Notations:}
\vspace{4 mm}

\noindent $mdv2_{(p,q)}(A)$ = $A$ $mod(2^p)div(2^q)$. ( A being an integer, mod modulo operation , and \\*
\noindent \hspace{10 mm} div integer division ).\\*

\noindent $\parallel A \parallel$ : the size in bits of A. \\*\\*

\subsection{ ModDiv2Inv and Subset Sum problem  : }
\vspace{10 mm} 

\newtheorem*{mydef1}{Definition 1}
\begin{mydef1}

ModDiv2Inv is a problem consisting on inverting 
the function below :
\\*

\noindent $F(X) =mdv2_{(p,q)}(A\times X)$.\\*\\*  A,  p and q are known integers where A is pseudorandom, $\parallel A \parallel$ = p, $\parallel X \parallel$ = q  and $p > 2 \times q $.
\end{mydef1}
\vspace{4 mm}

\newtheorem*{mydef2}{Definition 2}
\begin{mydef2}
Given target integer $T$ and a set of positiv integers $S$= \{ $s_1,s_2,...,s_n$ \} \\* Subset sum problem asks to find a subset of $S$ that sums up to $T$.
\\*

\noindent Or find an n bit integer $X : x_1x_2...x_n$ such as : 
$ T = \sum\limits_{i=1}^{n}  x_i \times s_i $.     

\end{mydef2}
\vspace{4 mm}

\newtheorem*{mydef3}{Definition 3}
\begin{mydef3}
Given integers $m$, $T$ and a set of positiv integers $S$ = \{ ${s_1,s_2,...,s_n}$ \} where the first element $s_1$ is pseudorandom and m bit, subsequent elements fullfill
the following relation $s_i = (2 \times s_{i-1}) mod (2^m) + r_i $,
$ r_i $ being pseudorandom in \{0,1\}.\\*

\noindent LS2R Subset sum problem ( LS2R\_SSP ) asks to find a subset of $S$ that sums up to $T$.
\vspace{4 mm}

\end{mydef3}
\newtheorem*{theo}{Theoreme 1}
\begin{theo}

\noindent ModDiv2Inv is equivalent to LS2R\_SSP.

\end{theo}

\noindent Proof. In what follows, we will show that ModDiv2Inv and LS2R\_SSP are reducible to each other proving then Theoreme 1. \\*

\newpage
 
\noindent\underline{\bf{1 - Reduction of ModDiv2Inv to LS2R\_SSP :}}\\* 
\vspace{4 mm}

\noindent Let $A$, $X$ and $Y$ be integers. $X : x_1x_2...x_q$ is q bits ( $x_{i=1 \rightarrow q}$ are in \{0,1\} ).\\*

$ Y = A \times X = A \times \sum\limits_{i=1}^{q}  x_i \times 2^{i-1} = \sum\limits_{i=1}^{q} x_i \times A \times 2^{i-1} $.\\*

\noindent Notice if $ Y \equiv 0 \hspace{2 mm} Mod(A) $, performing division 
$ Y / A $ is basically finding a subset of \\* 
\noindent set $S = {A \times 2^{0},A \times 2^{1},...,A \times 2^{q-1}}$ that sums to $Y$.\\*

\noindent Now let $S$ be a set of integers \{ $s_1,s_2,...,s_{q}$ \} where 
$s_i = mdv2_{(p,q)}(A \times 2^{i-1})$\\*

\begin{figure}[H]

\caption{}
\raggedleft
\includegraphics[scale=1.00]{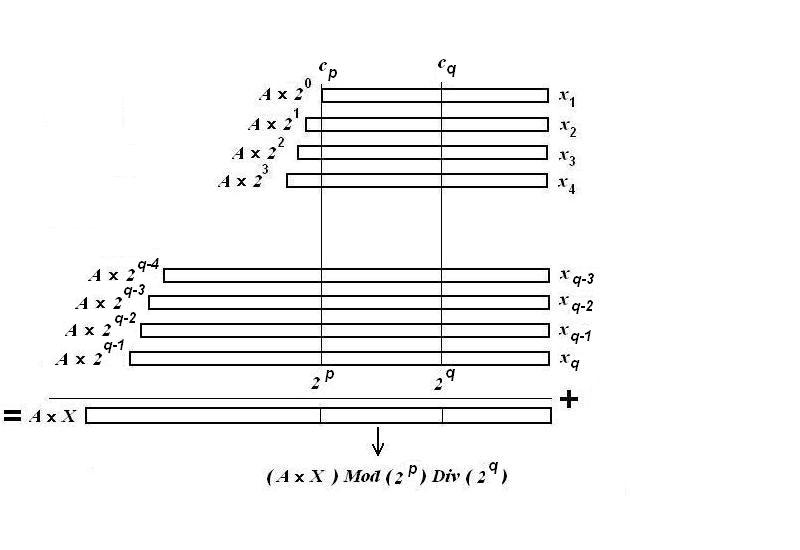}

\end{figure}

\noindent From Figure 1, one can see that  
$ \sum\limits_{i=1}^{q}  x_i \times s_i  =mdv2_{(p,q)}(A\times X) + c_{p}\times 2^{p} -  c_{q} $\\*

\noindent Where $c_{q}$ is the qth carry of multiplication $(A\times X)$ and $c_{p}$ is the pth carry of sum $ \sum\limits_{i=1}^{q}  (x_i \times s_i)$.\\* 
\noindent Both $c_{p}$ and $c_{q}$ can not be greater than $q$, the size in bits of $X$.\\*

\noindent To solve ModDiv2Inv or invert $F(X) =mdv2_{(p,q)}(A\times X)$, one had to find 
\noindent a q bit integer $X : x_1x_2...x_q$ such as  
$ F(X) + c_{p}\times 2^{p} -  c_{q}  = \sum\limits_{i=1}^{q}  x_i \times s_i $. $c_{p}$ and $c_{q}$ varie between $0$ and $q$.\\*

\noindent Or solve $q^2$ subset sum problems that asks to find
subsets of $S$ that sums up to targets \\*
 $F(X) + c_{p}\times 2^{p} -  c_{q}$. one of these subsets will
 correspond to X.
  \\*

\noindent Recall, elements of $S$ are \{ ${s_1,s_2,...,s_{q}}$ \} (
$s_i = mdv2_{(p,q)}(A \times 2^{i-1})$). \\* 

\noindent Now let's suppose that $s_1 = mdv2_{(p,q)}(A) = v_1 mod (2^m) $ where $ m = p - q $. \\*

\noindent Figure 1 in mind we can easily observe that : \\* 

$s_i = ( 2 \times s_{i-1} ) mod (2^m) + mdv2_{(i+1,i)}(A) $.\\*

$r_i = mdv2_{(i+1,i)}(A)$.\\*   

\noindent A is pseudorandom then both $s_1$ and $r_i$   
which is in \{ 0,1 \}, are also pseudorandom, meaning that
the $q^2$ subset sum problems to be solved are instances of LS2R\_SSP.\\*

\noindent Following is an algorithm solver for ModDiv2Inv.\\*

\noindent\line(1,0){436}\\*
\noindent \textbf{Inputs :} $S$ = \{ $s_1,s_2,...,s_{q}$ \} ( 
$s_{i=1 \rightarrow q} = mdv2_{(p,q)}(A \times 2^{i-1})$ ) ,   $Y$ = $mdv2_{(p,q)}(A \times X) $ )\\*
\noindent\underline{\textbf{Output :} $X$ \hspace{130 mm}}\\*

\noindent 1 : \textbf{For} i = 0 \textbf{To} q\\*
\noindent 2 :\hspace{5 mm}  \textbf{For} j = 0 \textbf{To} q\\*
\noindent 3 :\hspace{10 mm}  $X_0 = LS2R\_SSP\_Solver( S ,  Y + i \times 2^{p} -  j )$\\*
\noindent 4 :\hspace{10 mm}  \textbf{If} ($Y$ = $mdv2_{(p,q)}(A \times X_0))$  Output $X_0$\\*
\noindent 5 :\hspace{5 mm} \textbf{End For}\\*
\noindent 6 : \textbf{End For}\\*
\noindent\line(1,0){436}\\*

\noindent Note, this algorithm efficiency depends on the hardness of subset sum instances : \\*

\noindent [ $ S =$ \{ ${s_1,s_2,...,s_{q}}$ \} ( 
$s_{i=1 \rightarrow q} = mdv2_{(p,q)}(A \times 2^{i-1})$ ) , $ Y + i \times 2^{p} -  j $ ], $i$ and $j$ varie from $0$ to $q$.  \\*  

\newpage

\noindent \underline{\bf{ 2 - Reduction of LS2R\_SSP to ModDiv2Inv  : }} \\*

\vspace{4 mm}

\noindent Let T be an integer and $S$ be a set of integers  \{$s_1,s_2,...,s_n$\} \\*

\noindent $s_1$ is m bit and $s_{i=2 \rightarrow n} = (2 \times s_{i-1}) mod (2^m) + r_i $, $ r_i $ being pseudorandom in \{ 0,1 \}.  \\*

\noindent To reduce LS2R-SSP to ModDiv2Inv.\\*

\noindent We had to find
integers A, p, and q such as a n bits integer $X : x_1x_2...x_n$ exists and satifies : \\*

a - $ T = \sum\limits_{i=1}^{n}  x_i \times s_i $ \\*

b - $T + c_q = mdv2_{(p,q)}(A \times X)$\\*

\noindent $c_q$ being the qth carry of multiplication $A \times X$.\\* 

\noindent Bearing in mind Figure 1, it is easy to see that $q = n$, $p = m + q$.\\*

\noindent A can be computed by the following algorithm : \\*

\noindent 1 : $A = s_1$\\*
\noindent 2 :  \textbf{For} i = 2 \textbf{To} m\\*
\noindent 3 :\hspace{5 mm}$A = 2 \times A + s_{i-1} mod (2)$\\*
\noindent 4 : \textbf{End For}\\*

\noindent LS2R-SSP can be solved by the following algorithm : \\*

\noindent\line(1,0){436}\\*
\noindent \textbf{Inputs :} $S$ = [ ${s_1,s_2,...,s_{q}}$ ] ( 
 $s_i = (2 \times s_{i-1}) mod (2^m) + r_i $ ) ,
   A, p, q\\*
\noindent\underline{\textbf{Output :} $SS$ a Subset of $S$ \hspace{105 mm}}\\*

\noindent 1  \hspace{2 mm}:  $T_0 = 0$\\*
\noindent 2  \hspace{2 mm}: \textbf{For}  i = 0 \textbf{To} q\\*
\noindent 3  \hspace{2 mm}:\hspace{5 mm} $X = invert(T+i = mdv2_{(p,q)}(A \times X) )$\\*
\noindent 4  \hspace{2 mm}:\hspace{5 mm} \textbf{For} j = 0  \textbf{To} q\\*
\noindent 5  \hspace{2 mm}:\hspace{10 mm} $x_j = mdv2_{(j+1,j)}(X) )$\\*
\noindent 6  \hspace{2 mm}:\hspace{10 mm} \textbf{IF} ( $x_j = 1$ )\\*
\noindent 7  \hspace{2 mm}:\hspace{15 mm} Put $s_j$ into set $SS$\\*
\noindent 8  \hspace{2 mm}:\hspace{15 mm} $T_0 = T_0 + s_j$ \\*
\noindent 9  \hspace{2 mm}:\hspace{10 mm} \textbf{End IF} \\*
\noindent 10 :\hspace{5 mm} \textbf{End For}\\*
\noindent 11 :\hspace{5 mm} \textbf{IF} ( $T_0 = T+i$ ) Output $SS$\\*
\noindent 12 : \textbf{End For}\\*
\noindent\line(1,0){436}\\*

\noindent Note in line 3, $i = c_q $ is a possible qth carry of multiplication $A \times X$ ( figure 1 ).\\*

\subsection{ModDiv2Inv Hardness  : }
\vspace{6 mm}

\newtheorem*{mydef4}{Definition 4}
\begin{mydef4}

Let a Subset sum problem $P$, consisting on finding a subset of a set of positiv integers $S$ = \{$s_1,s_2,...,s_{n}$ \} that sums up to target $T$.
The density $D(P)$ of $P$ is defined as : \\*

\fontsize{0.4cm}{1pt}$D(P)$ =\fontsize{0.5cm}{1pt} $\frac{n}{Max \hspace{1 mm} Log_2 \hspace{1 mm} (s_i) \hspace{1 mm} : \hspace{3 mm} 1 \leq i \leq n }$

\end{mydef4}
\vspace{10 mm}

\noindent It has been found that subset sum problems whose densities are below 0.9408, are solvable by a lattice oracle, those whose densities are above 1, are solvable by dynamic programming. \\*

\noindent  It is assumed that the hardest subset problems are those whose densities are between 0.9408 and 1 [4][5][6].\\*

\noindent We assume that it is also the case for LS2R-SSP.\\*

\noindent In order to be equivalent to the hardest LS2R-SSP instances, parameters of a ModDiv2Inv should fullfill following conditions :\\*

1 - The bits composing $A$ are generated randomly.\\*

2 - $0.9408 < q / (p - q) \leq 1$. \\*

\noindent q/( p - q ) is the underlying LS2R-SSP density.\\*

\noindent X is q bits ( Figure 1 ). Condition 2, follows from the fact that : \\*

Max($  \parallel s_{i=1 \rightarrow q} = mdv2_{(p,q)}(A \times 2^{i-1}) \parallel$) = $p-q$.

\newpage

\section{ModDiv2Kex : } 
\vspace{4 mm}
\subsection{Algorithm : }
\vspace{4 mm}
\subsubsection{Public parameters : }
\vspace{4 mm}
\noindent Integers A, p, q and S. \\*

\noindent Real number D. \\*

\noindent A is pseudorandom and $\parallel A \parallel = p$.\\*

\noindent $D = q / (p - q)$, should be included in [ 0.9408 , 1 ].\\*

\noindent S is exchanged key size and equal to $ p - (2 \times q) $.\\*

\noindent Below is a table containing values for p and q corresponding to exchanged key size 128 and D from 0.95 to 0.99. \\*

\begin{tabular}{|l|l|l|l|}
\hline

$D = q/q+p$&$S$&$q=S\times D/(1-D)$&$p =2q+S$\\

\hline
0.95&128&2432&4992\\
\hline
0.96&128&3072&6272\\
\hline
0.97&128&4139&8046\\
\hline
0.98&128&6272&12672\\
\hline
0.99&128&12672&25472\\
\hline

\end{tabular}

\subsubsection{Private Computations : }
\vspace{4 mm}

\noindent - Alice generates randomly a q bit number X, and calculates $U = mdv2_{(p,q)}(A\times X)$.\\*

\noindent - Bob generates randomly a q bit number Y, and calculates $V = mdv2_{(p,q)}(A\times Y)$.\\*

\subsubsection{Public Exchange of Values : }
\vspace{4 mm}

\noindent - Alice sends $U$ to Bob.\\*

\noindent - Bob sends $V$ to Alice.\\*
 
\subsubsection{Further Private Computations : }
\vspace{4 mm}

\noindent - Alice calculates $Wa = mdv2_{(p,p-q)}(X\times V)$.\\*

\noindent - Bob calculates $Wb = mdv2_{(p,p-q)}(Y\times U)$.\\*

\noindent Bob and Alice know that : \\*

 $Wa$ = $Wb$ or $\mid Wa - Wb \mid$ = 1.\\*

\subsection{Proof of correctness : }
\vspace{4 mm}

\newtheorem*{mydef5}{Lemma 1}
\begin{mydef5}
\noindent A, p and q are integers  . \\*

\noindent if $p > q$ $\Rightarrow$ $mdv2_{(p,q)}(A\times 2^q)$ = $A$ $mod(2^{p-q})$. \\*

\end{mydef5}

\noindent \underline{Proof :} \\* 

$mdv2_{(p,q)}(A\times 2^q)$ = $(A\times 2^q)mod(2^{p})div(2^q)$. \\*

\noindent Observe lest significant $q$ bits of $N =(A\times 2^q)$ $mod(2^p)$ are zeros whereas its most significant $p-q$ bits are the lest significant $p-q$ bits of $A$, dividing then $N$ by $(2^q)$ implies : \\*

 $(A\times 2^q)mod(2^{p})div(2^q)$ = $A$ $mod(2^{p-q})$ = $mdv2_{(p,q)}(A\times 2^q)$  . \\*

\newtheorem*{mydef6}{Theoreme 2}
\begin{mydef6}
\noindent A, X, Y, p and q are integers where $p > q$, $\parallel A \parallel = p$, $\parallel X \parallel = \parallel Y \parallel = q$.\\*

\noindent $Wa$ = $mdv2_{(p-q,q)}(X \times mdv_{(p,q)}(A\times Y))$.\\*

\noindent $Wb$ = $mdv2_{(p-q,q)}(Y \times mdv_{(p,q)}(A\times X))$.\\*

\noindent There is two possibilities : \\*

1 - $Wa$ = $Wb$. \\*

2 - $\mid Wa - Wb \mid$ = 1.\\* 
\end{mydef6}

\newpage

\noindent \underline{Proof :} \\*

\noindent Let $H1$ and $H2$ be integers such as : \\*

$U_1 = mdv2_{(p,q)}(A\times X) \times 2^q = (A \times X - H1 ) mod(2^p)$. \\*

$V_1 = mdv2_{(p,q)}(A\times Y) \times 2^q = (A \times Y - H2 ) mod(2^p)$. \\*

\noindent The fact that the lest significant q bits of $U_1$ and $V_1$ are zeroes implies $\parallel H_1
\parallel = \parallel H_2 \parallel = q $. \\*

\noindent Let's calculate : \\*

$Wa_1$ = $(X \times V_1)mod(2^p)$ = $((X \times Y \times A) - (X \times H_2) )mod(2^p)$ (1)\\*

$Wb_1$ = $(Y \times U_1)mod(2^p)$ = $((Y \times X \times A) - (Y \times H_1) )mod(2^p)$ (2)\\*

\noindent  $\parallel X
\parallel = \parallel Y \parallel =\parallel H_1
\parallel = \parallel H_2 \parallel = q $ implies 
$\parallel X \times H_2
\parallel = \parallel Y \times H_1 \parallel  = 2 \times q $, we have then : \\*

$Wa_1div(2^{2\times q})$ = $(X \times Y \times A)mod(2^p)div(2^{2\times q}) - E_a$\\*

$Wb_1div(2^{2\times q})$ = $(Y \times X \times A)mod(2^p)div(2^{2\times q}) - E_b$\\*

\noindent where $E_a$ and $E_b$ are rescpectively the $2\times q$'th borrows of binary substractions (1) and (2) .\\*

\noindent $E_a$ and $E_b$ being bits, they can have then for values 0 or 1 implying :\\*

\noindent if $E_a$ = $E_b$ we have $Wa_1div(2^{2\times q})$ = 
$Wb_1div(2^{2\times q})$.\\*

\noindent if $\mid E_a-E_b \mid = 1 $ we have $\mid Wa_1div(2^{2\times q})- Wb_1div(2^{2\times q}) \mid = 1$.\\*

\noindent Now we'll show that :\\*

$ Wa = Wa_1div(2^{2\times q})$ and $ Wb = Wb_1div(2^{2\times q})$\\*

\noindent ending thus theoreme's proof . \\*

$Wa_1$ = $(X \times V_1)mod(2^p)$ = $(X \times mdv2_{(p,q)}(A\times Y) \times 2^q)mod(2^p)$\\*

$Wa_1div(2^{q})$ = $(X \times V_1)mod(2^p)div(2^{q})$ = $(X \times mdv2_{(p,q)}(A\times Y) \times 2^q)mod(2^p)div(2^{q})$\\*

\newpage

\noindent Applying Lemma 1, we get :\\* 

$Wa_1div(2^{q})$ = $(X \times V_1)mod(2^p)div(2^{q})$ = $(X \times mdv_{(p,q)}(A\times Y))mod(2^{p-q})$\\*

$Wa_1div(2^{2 \times q})$  = $(X \times mdv2_{(p,q)}(A\times Y))mod(2^{p-q})div(2^{q})$\\*

$Wa_1div(2^{2 \times q})$  = $mdv2_{(p-q,q)}(X \times mdv_{(p,q)}(A\times Y))$\\*

$Wa_1div(2^{2 \times q})$  = $Wa$\\*

\noindent by the same way we can prove : \\*

$Wb_1div(2^{2 \times q})$  = $Wb$\\*

\noindent Observe, $S = Max(\parallel Wa \parallel)$ = $Max(\parallel Wb \parallel)$ = $p - (2 \times q)$. \\*

\noindent A python implementation of ModDiv2Kex with  parameters q = 2432 and S = 128 is provided in Appendix A.\\*

\subsection{ModDiv2Kex security : } 
\vspace{4 mm}
\newtheorem*{mydef7}{Definition 5}
\begin{mydef7}

\noindent\ The computationnel ModDiv2 assumption states : \\*

\noindent\ Let G be a function defined as : \\*

$G_{A,p,q}(x) =mdv2_{(p,q)}(A\times x)$.\\*

\noindent\ A, p, q are integers, $\parallel A \parallel$ = $\parallel x \parallel$ = q  and  $0.9408 < q /(p-q) \leq 1$ .\\* 

\noindent\ Given $g_1 = G_{A,p,q}(x_1)$, $g_2 = G_{A,p,q}(x_2)$ ( $x_1$ and $x_2$ are unknown ) : \\*

\noindent\ Computing $k_1 = G_{g_2,p-q,q}(x_1)$ or $k_2 = G_{g_1,p-q,q}(x_2)$ is intractable. \\*

\end{mydef7}

\noindent ModDiv2Kex security is based on the difficulty of finding $X$ and $Y$, $Wa = mdv2_{(p-q,q)}(X \times V)$ and $Wb = mdv2_{(p-q,q)}(Y \times U)$ while knowing :\\*

\noindent $A$, $p$, $q$, $r$, $U = mdv2_{(p,q)}(X\times A)$ and $V = mdv2_{(p,q)}(Y\times A)$.\\*

\newpage

\noindent To be secure under the computationnel ModDiv2 assumption  :\\*

\noindent ModDivKex parameters $A$, $p$ and $q$ should satisfy the following below :\\*

1 - The bits composing $A$ are generated randomely.\\*

2 - $0.9408 < q /(p-q) \leq 1$ \\*

\noindent Summing it up, ModDiv2Kex will be unsecure, if one find how to solve efficiently subset sum problems which densities are between 0.9408 and 1 or proove that the computationnel ModDiv2 assumption is false
.\\*

\section{Implementation and efficiency :  } 
\vspace{2 mm}

\begin{figure}[H]

\caption{}
\raggedleft
\includegraphics[scale=1.00]{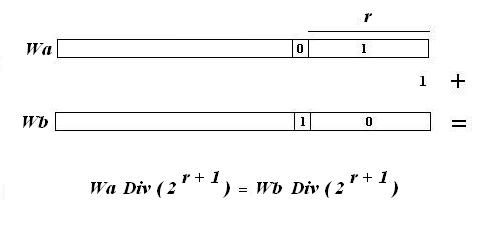}

\end{figure} 

\noindent In comparaison to Diffie-Hellman in the multiplicatif group and RSA that have time complexities of O($n^3$), 
the key exchange algorithm presented in this paper can be realised by a multiplication circuit where some leftmost and righmost output bits are discarded, meaning that it has a time complexity of O($n^2$) and with the same securiy parameter $n$, it can be $O(n)$ time faster than Diffie-Hellman in the multiplicatif group and RSA algorithms.\\*

\noindent But there is one drawback, Alice or Bob don't get  precisely the same value : they only know that  $Wa$ = $Wb$ or $\mid Wa - Wb \mid$ = 1.\\*

\noindent Meaning that if Alice encrypt a message M with key $Wa$ and sends corresponding cipher text C to Bob. To get M, he should decrypt C with $Wb$, $Wb+1$ and $Wb-1$ :  and gets 3 plausible plain texts. To decide which one is correct, Alice should hash M and join the computed digest as a header to M before encryption. Bob can then decide which plain text is correct by hashing obtained plaine text and compare it to joined hash value : Decryption can be then three times slower than encryption.\\*

\noindent Now Lets suppose that computed values $Wa$ and $Wb$ are uniform so that the probability that their least r significant bits are ones is $(1/2)^{-r}$.\\*

\noindent To get M, Bob can perform only one decryption if he and Alice agreed on r. Alice should then encrypt with $Wa$ $Div( 2^{r})$, Bob should decrypt with $Wb$  $Div(2^{r})$ (Figure 2).\\*

\noindent We have experimentaly observed that $Pr[Wa = Wb] = 2/3$, implying : \\*

$Pr[$ $Wa$ $Div( 2^{r}) = Wb$ $Div( 2^{r})$ $] = (1/3) \times 1/2^{-(r-1)}$.\\*

\noindent Bob can then decrypt C only one time, but the price going with it is r bits security and a probability  of $(1/3) \times 1/2^{-(r-1)}$ to get the right plain text. \\*\\*

\newpage

\section{Conclusion and future work:} ~ \\*

\noindent In this paper we have analayzed ModDiv2Inv a problem consisting on inverting the function below : \\*

$F(X) =(A\times X)Mod(2^p)Div(2^q)$ .\\* 

\noindent Mod is modulo operation, Div is integer division operation, A, p and q are known integers where A is pseudorandom and p bit, X is q bit and $p > 2\times q$.\\*

\noindent We have evaluated ModDiv2Inv's hardness by reducing it to
subset sum problem, a well studied np complete problem,
and found that the hardest ModDiv2Inv instances satisfy
the following conditions  :\\*

1 - The bits composing $A$ are generated randomly.\\*

2 - $0.9408 < q /(p-q) \leq 1$ \\*

\noindent We have presented ModDiv2Kex a new public key exchange algorithm based on ModDiv2Inv.\\*

\noindent Then introduced a computationnel assumption related to the hardest instances of ModDiv2Inv, and showed what conditions ModDiv2Kex parameters should fullfill in order to be based on said assumption.\\*

\noindent ModDiv2Kex is very efficent compared to Diffie Hellman and RSA cryptosystems,
furthermore the fact that it is not based on classical problems namely factoring and discret logs over multiplicative groups or elliptic curves makes it elligible to postquantum cryptography [3][6].\\*

\noindent One can construct public key encryption and digital signature based on ModDiv2Inv problem, and it is also quite possible to device symmetric key algorithms based on
it, namely hash functions and pseudo random numbers generators.\\*

\noindent in the future we would study and construct public key cryptosystems based on the difficulty to invert $F(X) =(A\times X)Mod(P)Div(Q)$  
while P and Q being integers.

\newpage

\section{Appendix A :} ~ \\*

\noindent Following python script is a " practical " proof of correctness of key exchange algorithm presented in this paper. (pycrypto library is needed )\\*

\noindent ========================================================\\*
\noindent import sys \\*
\noindent from Crypto.Util.number import getRandomNBitInteger \\*

\noindent def ModDiv(A,B,C) :\\*
\indent   return (A \% B ) // C \\*

\noindent l = 4992 \\*
\noindent m = 2432 \\*
\noindent p = 4992 \\*
\noindent q = 2432 \\*
\noindent r = 28 \\*

\noindent " Size in bits of public pararameter A is l " \\*
\noindent  A = getRandomNBitInteger(l,randfunc=None) \\*\\*

\noindent " Size in bits of private parameters X and Y is m "\\*
\noindent   X = getRandomNBitInteger(m,randfunc=None)\\*
\noindent   Y = getRandomNBitInteger(m,randfunc=None)\\*
 
\noindent  M = pow(2,p) \\* 
\noindent  M1 = pow(2,p-q) \\*
\noindent  D = pow(2,q) \\*
\noindent  D1 = pow(2,m+r) \\*

\noindent " If r = 0,  In 30 \% percents, the keys computed by Alice and Bob are not identical : " \\*
\noindent " $Wa = Wb \pm 1$, this is due to bit carry propagation, if r is increased by one " \\*
\noindent " the probability that Wa is diffrent to Wb is devided by two.  " \\*

\noindent U =  ModDiv(A*X,M,D) 

\noindent V = ModDiv(A*Y,M,D)
   
\noindent Wa = ModDiv(U*Y,M1,D1)

\noindent Wb = ModDiv(V*X,M1,D1) \\*

\noindent print("")  \\*
\noindent print("Public Parameters :")  \\*
\noindent print("===================")  \\*
\noindent print("")   \\*                  \\*
\noindent print("Public Parameter l = \%d" \%l)  \\*
\noindent  print("") \\*
\noindent print("Public Parameter m = \%d" \%m)  \\* 
\noindent  print("") \\*
\noindent print("Public Parameter p = \%d" \%p)  \\* 
\noindent  print("") \\*
\noindent print("Public Parameter q = \%d" \%q)  \\*
\noindent  print("") \\* 
\noindent print("Public Parameter r = \%d" \%r)  \\* 
\noindent  print("") \\*
\noindent print("Public Parameter A = \%d" \%A)  \\*
\noindent  print("") \\*
\noindent print("")   \\*
   
\noindent print("Private Parameters :") \\*
\noindent print("====================") \\*
\noindent print("") \\*           

\noindent print("Alice Private Parameter X = \%d" \%X)\\*
\noindent  print("") \\*
\noindent print("Bob Private Parameter   Y = \%d" \%Y)\\*
   
\noindent print("") \\*\\*\\* 

\noindent print("Shared Parameters :") \\*
\noindent print("=============") \\*
\noindent print("") \\*        

\noindent print("Parameter shared with Bob by Alice U = \%d" \%U)\\*
\noindent  print("") \\*
\noindent print("Parameter shared with Alice by Bob V = \%d" \%V)\\*
   
\noindent print("") \\* 

\noindent print("Exchanged Secret Key :") \\*
\noindent print("======================") \\*
\noindent print("") \\*

\noindent print("Secret key computed by Alice Wa = \%d" \%Wa) \\*
\noindent  print("") \\*
\noindent print("Secret key computed by Bob  Wb = \%d" \%Wb) \\*

\noindent  print("") \\*

\noindent sys.exit \\*
\noindent ========================================================\\*

\noindent You can download this script from : $https://github.com/Crypticator/ModDiv/blob/master/Kex1.py$ \\*\\*

\end{document}